\documentclass{article}
\usepackage[utf8]{inputenc}
\usepackage{amsmath}
\usepackage{amssymb}
\usepackage{listings}
\usepackage{graphicx}
\usepackage{multirow}
\usepackage{booktabs}
\usepackage{subcaption}
\usepackage[font=small]{caption}
\usepackage{dsfont}
\usepackage{xurl}
\usepackage{xcolor}
\usepackage[colorlinks=true, linkcolor=blue, citecolor=blue, urlcolor=blue]{hyperref}
\usepackage{natbib}
\usepackage[font=small]{caption}

\title{Scalable Dirichlet Process Mixture Models with Unknown Concentration and Adaptive Covariance for High-Dimensional Clustering Applied to Leukemia Transcriptomics}
\author{
        Annesh Pal, Aguirre Mimoun, 
        Rodolphe Thi{\'e}baut and Boris P. Hejblum$^{*}$
}
\date{}

\begin{document}

\maketitle
\begin{center}
\begin{minipage}{0.9\textwidth}
\small

%\textcolor{red}{BORIS: Nice abstract but could be improved : really focus on the strength of DPMM for clustering, ie estimating unknown number of clusters, which requires a varying concentration alpha and has been proven consistent, but currently lacks VI implementation with varying alpha and unknown and does not scale.}
\section*{Abstract}
Dirichlet process mixture models (DPMM) provide a flexible framework for clustering complex biological data without pre-specifying the number of clusters. Although Markov Chain Monte Carlo (MCMC) methods have bridged the gap between theory and application for such models, they scale poorly to high-dimensional data (e.g. omics data) and suffer from slow and difficult convergence. Variational Inference (VI) provides a faster alternative, but lacks implementation without overly simplifying assumptions such as mean field independence and known covariance. We propose a novel methodology that performs scalable inference of DPMM for clustering using collapsed VI, ensuring coherent and data-driven estimation of the cluster number in a fully Bayesian scheme. To accommodate high-dimensional data, we introduce cluster-specific adaptive covariance modeling with sparsity-inducing priors that improves flexibility while mitigating variance–covariance coupling and over-parameterization. The resulting algorithm scales efficiently with dimension and sample size, substantially reducing computation time compared to state-of-the-art MCMC slice sampler. Realistic simulation studies under Gaussian and negative-binomial settings demonstrate accurate recovery of cluster structure and robust performance across varying signal-to-noise regimes, supported by sensitivity analyses. Application on publicly available leukemia transcriptomic dataset comprising 72 samples and 2,194 gene expressions successfully recovers established subtypes while revealing biological plasticity of one Mixed Lineage Leukemia outlier sample. This theoretically coherent solution for high-dimensional non parametric Bayesian clustering is implemented in the companion R package \texttt{vimixr}.
\medskip

\noindent\textbf{Keywords:} Bayesian non-parametrics; Dirichlet process mixture model; Gene expression data; High-dimensional clustering; Variational inference.
\end{minipage}
\end{center}

\section{Introduction}
\label{sec:1}

Clustering is an exploratory data analysis technique to group the data based on measured or estimated inherent similarities \citep{jain2010data}. Since the early emergence of $K$-means from 1956 onward \citep{steinhaus1956division, mcqueen1967some}, clustering has become increasingly prevalent in complex biological data analysis, with diverse approaches including density based \citep{ester1996density, campello2013density}, graph based \citep{ertoz2002new, traag2019louvain} and probabilistic based approaches \citep{bock1996probabilistic, bouveyron2007high} %for better understanding of such high dimensional data. 
A key limitation, however, is that the number of clusters $K$ (or directly analogous parameters for density and graph based methods) must be preset to implement these approaches. 
%Even dimension reduction techniques like PCA and UMAP often fail to capture underlying structure for high-dimensional data and require strong domain-specific knowledge to select an appropriate $K$ \citep{wani2025comprehensive}. 
This is an ubiquitous problem in clustering, and
bayesian non-parametric mixture models (NPMM) have emerged as an efficient alternative to perform clustering based on probability distributions without pre-specifying $K$ \citep{gershman2012tutorial}.

\citet{ferguson1973bayesian} introduced the Dirichlet process $DP(\alpha, G_0)$ that underpins DP mixture models (DPMM) as a class of Bayesian NPMM for density estimation, clustering and regression \citep{muller2013bayesian}. DPMM leverage an infinite-dimensional parameter space thanks to a DP prior, yielding a posterior over the observations partitions space with variable $K$.
Although DPMM have been scrutinized for inconsistent posterior estimation of $K$  \citep{miller2013simple}, \citet{ascolani2023clustering} successfully demonstrated their consistency is guaranteed as long as the concentration parameter $\alpha$ is not fixed but data-driven \citep{escobar1995bayesian}.

DPMM often results in analytically intractable likelihoods and posteriors, making Markov Chain Monte Carlo (MCMC) methods a popular inference choice \citep{neal2000markov}. However, these methods exhibit slow and hard to assess convergence \citep{gershman2012tutorial}, especially for high dimensions with several hundreds or even thousands of features per sample and limited sample size. Variational inference (VI) approximates the true posterior $p(\cdot | data)$ with a tractable density function $q(\cdot)$ from a (known) family of probability distributions, while optimizing an information-theoretic criterion such as KL divergence \citep{blei2017variational}. Rooted in early estimation methods for graphical models \citep{jordan1999introduction}, VI has gained prominence for its computational efficiency and scalability, particularly in modern machine learning applications involving large datasets \citep{loya2025scalable}.

Standard DPMM implementation use a stick-breaking DP representation \citep{sethuraman1994constructive}, with hyper-priors for $\alpha$ and DP base distribution $G_0$. Mean-field VI is a widely adopted implementation approach that assumes factorized variational distributions with conjugate priors for model parameters \citep{blei2006variational, bishop2006pattern}. However, latent allocation variables in DPMM are inherently dependent through their shared mixture proportions across samples. Mean-field VI thus fails to capture the hierarchical relationship and results in inconsistency between cluster numbers estimated from $\alpha$ and the latent allocation variables (details in Supplementary material S4). \citet{kurihara2007collapsed} provided a collapsed VI framework by integrating out (collapsing) the stick-breaking proportions, thus defining a latent allocation  prior that depends only on $\alpha$ \eqref{Eq:1}. This integration, however, constructs a non-standard prior with no conjugate hyper-prior for $\alpha$, leaving collapsed VI without a working implementation that adaptively infers $\alpha$.

Beyond $\alpha$ that has a directly proportional effect on the inferred number of clusters \citep{teh2010dirichlet}, the model variance also significantly impacts the estimation of $K$ \citep{fruhwirth2010bayesian, hejblum2019sequential}, a practical challenge that is often ignored in DPMM implementations, e.g. \citet{blei2006variational}. The cluster variance particularly impacts high-dimensional data due to reduced signal-to-noise ratio, high variability and low inter-sample separation \citep{clarke2008properties}, where commonly used priors like Inverse-Wishart $IW$ receive criticisms for rigidity and strong variance-covariance coupling \citep{tokuda2025visualizing}. Factorized covariance representations (e.g. Cholesky decomposed factors) exhibit similar variance-covariance dependency due to shared factor elements by linear combinations. Alternative approaches include block diagonal matrices, latent factor models and element-wise distributions \citep{jing2024variance} to introduce conditional independence within the covariance matrices. Additionally, real-world data often deviate from Gaussian assumptions \citep{hejblum2019sequential}, and transcriptomic data exhibit overlapping clusters due to high dimensionality or biological factors introducing multimodality and high correlation \citep{clarke2008properties, yu2017clustering}. Cluster-specific distributions that regulate within-cluster spread as well as inter-cluster separability are therefore essential to obtaining robust results.

We propose a novel collapsed VI approach that jointly incorporates adaptive inference of $\alpha$ and model covariance within a DPMM. Combining Euler-Maclaurin and Taylor series approximations, we obtain a closed form variational distribution for $\alpha$ with a conjugate Gamma hyper-prior. After evaluating different covariance parameterizations (Table \ref{table:sigmaprior}) on Gaussian simulated data, we establish the model with sparsity inducing hyper-prior (namely, Sparse DPMM) as optimal (similar to \citet{jing2024variance}). Hyper-parameter sensitivity of Sparse DPMM is examined with high dimensional Negative Binomial simulations, and benchmarked against an MCMC slice sampler (\texttt{NPflow} from \citet{hejblum2019sequential}) for computation speed. We further apply Sparse DPMM to cluster acute lymphoblastic (ALL), mixed lineage (MLL) and acute myeloid (AML) leukemia sub-types based on gene expression dataset \citep{armstrong2002mll, de2008clustering} with $2,194$ genes and $72$ samples, and compare performance with $K$-means \citep{mcqueen1967some}, DBSCAN \citep{ester1996density}, HDBSCAN \citep{campello2013density}, shared nearest neighbourhood \citep{ertoz2002new}, Leiden \citep{traag2019louvain} and HDDC \citep{bouveyron2007high} techniques. Our Sparse DPMM outperforms the conventional methods for the Leukemia data, while identifying an additional sub-cluster that corresponds to the plastic nature of MLL sub-type. We provide the \texttt{vimixr} R-package implementing our approach and available from CRAN.

\section{Methods}
\label{sec:2}
\subsection{Problem set-up}
Consider $\mathbf{X}_n \in \mathbb{R}^d$ random variables sampled from an unknown mixture of distributions $F$
\begin{center}
        $\mathbf{X}_n|G \overset{\mathrm{i.i.d}}\sim F$  for $n = 1, \dots , N$
\end{center}
where $F(\mathbf{X}) = \int_{\Theta}f_{\eta}(\mathbf{X})G(d\eta)$ \citep{hejblum2019sequential}. We assume a multivariate Gaussian distribution for $f_{\eta}(\cdot)$ with mean $\mathbf{\mu}$ and covariance matrix $\mathbf{\Sigma}$ ($\eta = \{\mathbf{\mu}, \mathbf{\Sigma} \}$). $G$ is the unknown mixing distribution that characterizes the mixture components over the parameter space $\mathbf{\Theta}$, thus inducing clustering among $\mathbf{X}_n$. A DP prior on $G \sim DP(\alpha, G_0)$ with concentration parameter $\alpha$ and base distribution $G_0$ \citep{ferguson1973bayesian} gives us a non-parametric mixing distribution 
\begin{center}
    
        $G(\cdot) = \sum_{k=1}^{\infty}\pi_k\delta_{\eta_k}(\cdot)$ where $\eta_k \in G_0$ 
   
\end{center}
The mixing proportions $\pi_k$'s are drawn from a stick-breaking scheme \citep{sethuraman1994constructive}, which further introduce latent allocation variables $Z = \{z_n\}$ in the model that follows a categorical distribution with parameters $\{\pi_k\}$'s such that probability of $z_n = k$ is $\pi_k$ for every $n \in N$. The likelihood of $\mathbf{X}_n$ hence follows  
\begin{center}
    
        $\mathbf{X}_n \sim \prod_k MVN(\mathbf{\mu}_k, \mathbf{\Sigma}_k)^{\mathds{I}_{[z_n = k]}}$  
   
\end{center}
\citep{blei2006variational}. Our objective is to estimate $z_n$ as well as $\eta_k = \{\mu_k, \Sigma_k\}$ to quantitatively evaluate the partitioning of the parameter space $\mathbf{\Theta}$.

\subsection{Dirichlet process mixture model}
A hierarchical Bayesian model follows
\begin{center}
    $V_k \sim Beta (1, \alpha)$\\
    \smallskip
    $\pi_k := V_k \prod_{j<k} (1-V_j)$\\
    \smallskip
    $\mathbf{\mu}_k, \mathbf{\Sigma}_k \sim G_0 $\\
    \smallskip
    $z_n \sim Categorical\left(\{\pi_k\}\right)$\\
    \smallskip
    $\mathbf{x}_n|z_n \sim MVN(\mathbf{\mu}_{z_n}, \mathbf{\Sigma}_{z_n})$
\end{center}
This is an illustration of a Dirichlet process mixture model (DPMM), introduced by \citet{antoniak1974mixtures}.
The concentration parameter $\alpha$ has a direct impact on the posterior expectation of the number of non-empty clusters \citep{teh2010dirichlet}. \citet{escobar1995bayesian} have proposed a data augmentation scheme that leverages a Gamma hyper-prior $\alpha\sim Gamma(a,b)$ yielding a posterior distribution that adjust to the actual number of clusters observed in the data with consistency \citep{ascolani2023clustering}. With a conjugate choice for $G_0$, the model provides an estimate for the unknown parameters of interest.

\subsection{Choices for $G_0$}

The underlying structure of $\eta = \{\mathbf{\mu}, \mathbf{\Sigma} \}$ governs the distributional choice for $G_0$. We use a multivariate Gaussian $MVN(\mathbf{0}, \mathbf{\Sigma}_{\mathbf{\mu}})$ as the prior for $\mathbf{\mu}_k$. Table \ref{table:sigmaprior} presents several parameterizations for $\mathbf{\Sigma}$ (and our proposed associated prior distributions).
\begin{table}[!ht]
    \centering
    \resizebox{0.85\textwidth}{!}{
    \begin{tabular}{lllll}
        \toprule
        \textbf{Model} & \textbf{Assumption} & \textbf{Parameterization} & \textbf{Cluster-specific} & \textbf{Prior} \\
        \midrule
        $M1$ & Known & $\frac{1}{\sigma} I_{d \times d}$ & No & -- \smallskip\\
        $M2$ & Known & $\mathbf{\Sigma}$ & No & -- \medskip\\
        $M3$ & Unknown & $\frac{1}{\sigma} I_{d \times d}$ & No & $\sigma \sim \Gamma(g_1, g_2)$ \smallskip\\
        $M4$ & Unknown & $\mathbf{\Sigma}$ & No & $\mathbf{\Sigma} \sim IW(\nu_0, V_0)$ \smallskip\\
        $M5$ & Unknown & $\mathbf{\Sigma}^{-1} = \mathbf{LL}^t$ & No & $\mathbf{L}_{ij} \sim N(\mu_0, \sigma_0), \mathbf{L}_{ii}^2 \sim \Gamma(a_0, b_0) $ \medskip\\
        $M6$ & Unknown & $\mathbf{\Sigma}_k$ & Yes & $\mathbf{\Sigma}_k \sim IW(\nu_0, V_0)$ \smallskip\\
        $M7$ & Unknown & $\mathbf{\Sigma}_k$ & Yes & $\mathbf{\Sigma}_{k_{ij}}^{-1} \sim Lap(0,c_0), \mathbf{\Sigma}_{k_{ii}}^{-1} \sim \Gamma(a_0, b_0){\mathds{1}_{[\mathbf{\Sigma}_{k}^{-1} \in M^+]}}$ \smallskip\\
        $M8$ & Unknown & $\mathbf{\Sigma}_k$ & Yes & $\mathbf{\Sigma}_{k_{ij}}^{-1} \sim N(c_0, 10^{-6}), \mathbf{\Sigma}_{k_{ii}}^{-1} \sim \Gamma(a_0, b_0){\mathds{1}_{[\mathbf{\Sigma}_{k}^{-1} \in M^+]}}$ \\
        \bottomrule\\
    \end{tabular}
    }
    \caption{Model structure based on choice of covariance matrix $\mathbf{\Sigma}$; $\sigma$ is a positive scalar quantity, $\mathbf{L}$ is the Cholesky factorized lower triangular matrix for $\mathbf{\Sigma}^{-1} = \mathbf{LL}^t$, $\mathbf{M}^+$ represents the set of positive definite matrices \citep{wang2012bayesian} and $i,j \in {1, \dots, d}$}
    \label{table:sigmaprior}
\end{table}
We introduce gradual complexity in terms of covariance parameterization, starting from known variance ($M_1, M_2$). For unknown variance, global refers to a common covariance matrix $\mathbf{\Sigma}$ for all cluster ($M_3, M_4, M_5$), whereas cluster-specific defines each cluster with a unique covariance matrix $\mathbf{\Sigma}_k$, along with the mean vectors ($M_6, M_7, M_8$). The choices of prior distributions all maintain conjugacy, while considering hyper-parameters that establish weakly informative priors (see supplementary Materials S2). Of note, all eight parameterizations and prior choices guarantee that covariance matrices always remain semi-definite positive (see Supplementary materials S2).

\subsection{Posterior estimation using collapsed VI}

We derive a collapsed variational inference (CVI) approach, where we integrate out (or collapse) the stick-breaking parameters $\{V_k\}$, following \citet{kurihara2007collapsed}. This yields a probability distribution $p_0$ of the latent allocation variables $\{z_n\}$ depending only on the DP concentration parameter $\alpha$: 
\begin{equation}\label{Eq:1}
    \begin{split}
        p_0(z_n|\alpha) & = \prod_k \int_{V_k} p_0(z_n|V_k)p(V_k|\alpha)dV_k\\
        & = \prod_k \alpha \frac{\Gamma(\mathds{1}_{[z_n=k]}+1)\Gamma(\mathds{1}_{[z_n>k]}+\alpha)}{\Gamma(1 + \mathds{1}_{[z_n \geq k]}+ \alpha)}
    \end{split}
\end{equation}
where $\mathds{1}_{[\cdot]}$ represents the indicator function.\\
We can reconstruct our DPMM as:
\begin{center}
    $\alpha \sim Gamma(a,b)$\\
    \smallskip
    $\mathbf{\mu}_k, \mathbf{\Sigma}_k \sim G_0 $\\
    \smallskip
    $z_n \sim p_0(z_n|\alpha)$ \eqref{Eq:1}\\
    \smallskip
    $\mathbf{x}_n|z_n \sim MVN(\mathbf{\mu}_{z_n}, \mathbf{\Sigma}_{z_n})$.\\
\end{center}
The posterior distribution is approximated by a mean-field variational distribution for the DPMM parameters 
\begin{center}
    $q(\alpha, \{\mu_k\}, \{\Sigma_k\}, \{z_n\}) = q(\alpha)\prod_{k \leq K} q(\mathbf{\mu}_k)q(\mathbf{\Sigma}_k)\prod_n q(z_n)$
\end{center}
\citep{blei2017variational}. Although having theoretically ``infinite`` prior choices for $k$, truncated variational distributions with an upper limit $k=K$ enable practical implementation \citep{ishwaran2001gibbs}.\\
For posterior estimation, we use exponential families of distributions for $q(\cdot)$ and apply coordinate ascent algorithm to update the variational hyper-parameters \citep{bishop2006pattern}. It ensures guarantee of convergence and provides a closed form update $q^{*}(\cdot)$ that maximizes the Evidence lower bound (ELBO) (or equivalently minimizing the information-theoretic Kullbach-Leibler (KL) divergence)
\begin{equation}\label{Eq:2}
    \begin{split}
        q^{*}(\theta_i) & \propto \exp(\mathbb{E}_{q_{\theta^{-i}}}\left[\log\,p(\mathbf{X}, \mathbf{\theta})\right])\\
        & \propto \exp(\mathbb{E}_{q_{\theta^{-i}}}\left[\log\,p(\mathbf{X}| \mathbf{\theta}) + \log\,p(\mathbf{\theta})\right])
    \end{split}
\end{equation}
where $\theta_j \in \mathbf{\Theta} = \{\alpha, \{\mathbf{\mu}_k\}, \{\mathbf{\Sigma}_k\}, \{z_n\}\}$ and $\theta^{-j} = \mathbf{\Theta} \backslash \theta_j$ represents the set of all except the $j^{th}$ parameter \citep{blei2006variational, bishop2006pattern}. Appropriate prior as well as variational distribution choices facilitate the update of $q^*(\cdot)$ hyper-parameters. These updates are used to perform posterior inference for the unknown $\{z_n\}, \{\theta_i\}$, thus estimating data cluster as well as distributional partitions corresponding to the cluster allocations. The theoretical updates of the variational parameters involved are fully detailed in the Supplementary material S1-S3. One important innovation from our collapsed VI framework is updating $\alpha$ based on the number of non-empty clusters $t<K$ only, contrary to the mean-field VI implementation from \citet{blei2006variational} that considers all potential clusters up to the truncation $K$ (including empty ones). Thus, linking $Z$ directly to $\alpha$ (cf Eq: \eqref{Eq:1}) and together with our theoretical approximations, we can avoid the inconsistencies in cluster number estimates between $\alpha$ and $Z$ that plague \citet{blei2006variational} algorithm (additional details provided in Supplementary material S3 and S4).

Starting with an initial choice of latent allocation probabilities $q^\circ_{nk}$, the DPMM iteratively updates VI parameters based on the coordinate ascent algorithm (Eq: \eqref{Eq:2}) until ELBO converges \citep{blei2006variational}
\begin{equation}\label{Eq:3}
    \begin{split}
        \text{ELBO} & = \mathbb{E}_q\biggl[\log\,p(\mathbf{X}, Z, \{\mathbf{\mu}_k, \mathbf{\Sigma}_k\}, \alpha)\biggr] - \mathbb{E}_q\biggl[\log\,q(Z, \{\mathbf{\mu}_k, \mathbf{\Sigma}_k\}, \alpha)\biggr]\\
        & = \mathbb{E}_q\biggl[\log\,p(\mathbf{X} | Z, \{\mathbf{\mu}_k, \mathbf{\Sigma}_k\}, \alpha) + \log\,p(Z, \{\mathbf{\mu}_k, \mathbf{\Sigma}_k\}, \alpha)\biggr]\\
        & - \mathbb{E}_q\biggl[\log\,q(Z, \{\mathbf{\mu}_k, \mathbf{\Sigma}_k\}, \alpha)\biggr].
    \end{split}
\end{equation}
ELBO acts as a lower bound of log evidence $\log\,p(\mathbf{X})$ which minimizes reverse KL divergence between the target $p(\cdot)$ and variational $q(\cdot)$ distributions.

\subsection{Hyper-parameter tuning and model selection}\label{sec:2.5}

Initialization can significantly impact VI results, due to non-convexity of the ELBO \citep{blei2017variational}. This is mitigated by leveraging several random initializations and selecting the optimum based on variational log likelihood (VLL) term from Eq: \eqref{Eq:3}: $\mathbb{E}_q[\log\,p(\mathbf{X} | Z, \{\mathbf{\mu}_k, \mathbf{\Sigma}_k\}, \alpha)](= \mathbb{E}_q[\log\,p(\mathbf{X} | Z)]).$ VLL implicitly handles model complexity through expectation over $q(\cdot)$, while providing a numerical fit of the data based on updated variational distribution. It is therefore a good metric to select the best initial values, contrary to the ELBO which is over-influenced by the large number of parameters in high dimension (details in Supplementary material S5 and S6.2). 

For the selection of $G_0$ and associated hyper-parameters, we use mean pairwise $\mathbb{KL}$ divergence scores:
\begin{center}
    $KL_{mean}=\frac{1}{\binom{K}{2}}\sum_{i \neq j \in K} \frac{1}{2}\biggl[\mathbb{KL}(\mathbf{\theta}_i || \mathbf{\theta}_j) + \mathbb{KL}(\mathbf{\theta}_j || \mathbf{\theta}_i) \biggr]$,
\end{center}
for posterior cluster distributions $\mathbf{\theta}_i, \mathbf{\theta}_j$ where $i\neq j \in K$. This score can be interpreted as the distributional separability of the estimated clusters, and provides coherent validation for high dimensional data (instead of geometric counterparts like Silhouette coefficients, see details in Supplementary material S6.2). Standard model selection criteria like BIC fall short in this setting due to unknown cluster number \citep{liu2011parametric}, while the ELBO and VLL are both subjects to numerical artifacts under certain hyper-parameter regimes (e.g. large hyper-parameter values) and thus cannot properly discriminate extreme values from optimal ones.

\section{Results}
\label{sec:3}
\subsection{Numerical study}

\subsubsection{Effect of $G_0$ covariance structure and prior choices}
\label{Sec:3.1}

Posterior cluster number estimates depend on $\alpha$ as well as the structure and prior choice of the variance in $G_0$ \citep{hejblum2019sequential}. 
To compare model specifications from Table:\ref{table:sigmaprior}, we generated random $N=100$ multivariate Gaussian variables of dimension $d=2$ and grouped them into $K_{true}=2$ clusters, where both clusters share the same covariance (details in Supplementary material S10). Using our \texttt{vimixr} implementation, we estimated the posterior number of clusters $K_{post}$ as well as adjusted Rand index (ARI) scores over the generated data for $100$ simulation runs.
\begin{figure}[!ht]
    \centering
    \includegraphics[width=0.75\textwidth]{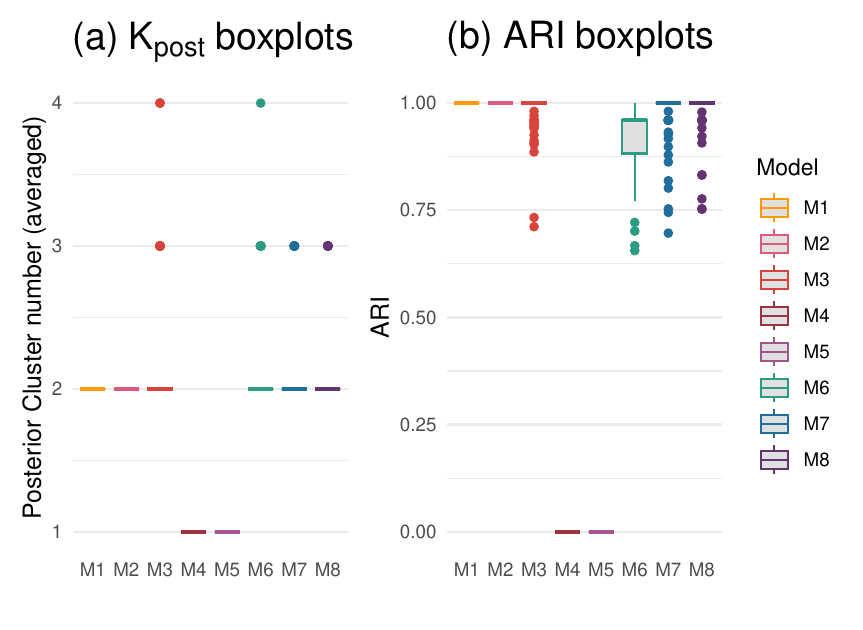}
    \caption{$K_{post}$ and ARI scores for different model choice ($K_{true}=2$)}
    \label{fig:1}
\end{figure}
In Figure \ref{fig:1}, fixed variance models $M_1$ (diagonal covariance) and $M_2$ (full covariance) establish the performance baseline, while cluster-specific structures ($M_6$-$M_8$) are a necessary complement to $\alpha$ in order to successfully recover the posterior number of non-empty clusters when the covariance is unknown.
In this latter case, global diagonal covariance $M3$ achieves good performances (but only because the data were indeed generated accordng to this model), while both full covariance $M4$ and Cholesky-decomposed covariance $M5$ yield significantly poor performance. $M3$ employs uncorrelated mixture distribution for all the variational clusters, thus reducing rigidity and variance-covariance coupling. Such global covariance structures ($M_3$- $M_5$) can prove inadequate for clusters with heterogeneous covariances \citep{biernacki2002assessing}. Among the latter structures, an IW prior ($M_6$) introduces rigidity \citep{tokuda2025visualizing}, while element-wise priors ($M_7$, $M_8$) decouple variance and covariance terms using conjugate Gamma distributions on the diagonal elements and either Laplace ($M_7$) or Normal ($M_8$) distributions for the off-diagonals (while ensuring semi-definiteness and positivity). The cluster-specific models recovered better than global models, even though the simulated data had cluster agnostic covariance. This setting was deployed to avoid model misspecification issues. However, as expected, the cluster-specific models shows better performance for data simulated using cluster-specific covariance (see Supplementary materials S10).

Compared to global covariance, cluster-specific covariance structures provide better results overall. For different prior choices of the cluster-specific covariance structure, we further illustrated the effect of dimensionality and more clusters by increasing dimension to $d = 100$ and calculating $K_{post}$ and ARI for different $K_{true} \in \{4, 6, 8, 10\}$.
\begin{figure}[!ht]
    \centering
    \includegraphics[width=0.75\textwidth]{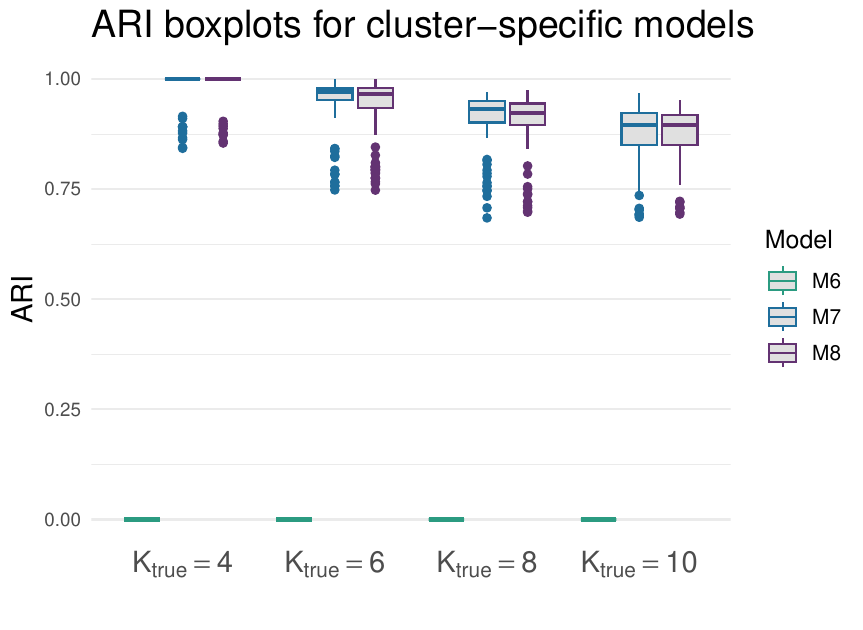}
    \caption{ARI for different cluster-specific models when $K_{true}$ is varying}
    \label{fig:2}
\end{figure}
Element-wise prior models show better adaptability with varying cluster number than $IW$ prior model in a high-dimensional setting (Figure \ref{fig:2}). $M_7$ and $M_8$ have comparable performances (with $M_7$ slightly better than $M_8$). This can be attributed to the fact that a Laplace distribution is equivalent to a marginalized Gaussian distribution with Exponential prior on the variance parameter. Owing to its sparsity inducing effect \citep{jing2024variance}, we call $M_7$ as the Sparse DPMM, and consider it to be the best possible choice among the models defined in Table \ref{table:sigmaprior}.

\subsubsection{Computational scalability}
We implemented our model framework and its VI estimation in the \texttt{vimixr} R-package (available from CRAN with all $8$ parameterizations from Table \ref{table:sigmaprior}). For the Sparse DPMM, computations scale linearly with the sample size and quadratically with the dimension (details in Supplementary material S7). Although convergence is often reached with fewer iterations in higher dimensions, the cost for a given $K$ is slightly higher due to computing larger $d$-dimensional precision matrices. We compared computing times with the slice-sampler MCMC algorithm implemented in the \texttt{NPflow} R package \citep{hejblum2019sequential} for $N = 100,  d = 100$ Gaussian data with $K=2$. With $1000$ MCMC iterations, the slice sampler often has not converged yet, while our Sparse DPMM systematically reaches convergence in timings that are three orders of magnitude less (details in Supplementary material S7).

\subsection{Application to cancer sub-type estimation}
\begin{figure}[!ht]
  \centering
  \includegraphics[width=0.75\textwidth]{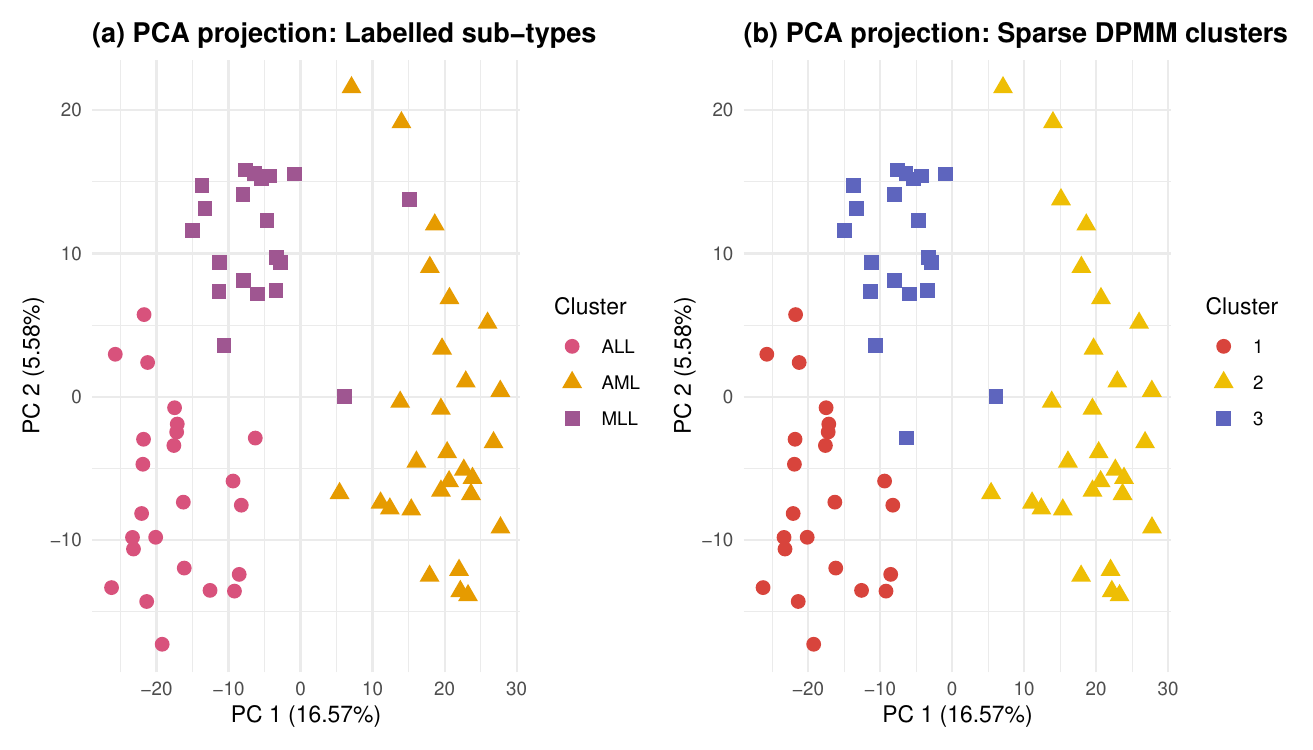}
  \caption{PCA projection on the first $2$ principal components for (a) labelled Leukemia sub-types based on $2194$ genes and (b) Sparse DPMM cluster estimates obtained using strong hyper-parameters}
  \label{fig:3}
\end{figure}

\begin{figure}[!ht]
  \centering

  \begin{subfigure}[b]{0.85\textwidth}
    \includegraphics[width=\textwidth]{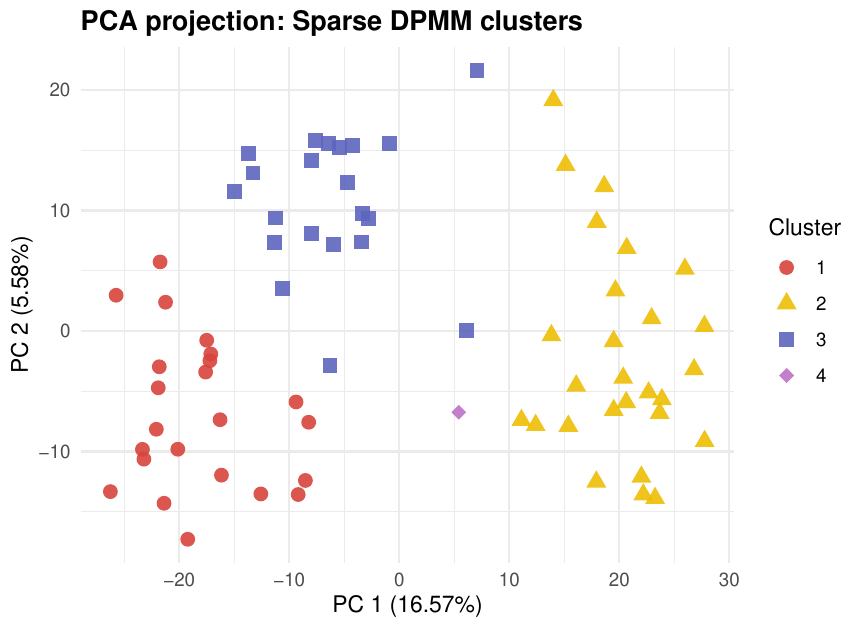}
    \caption{Sparse DPMM clusters using weaker hyper-prarameters}
    \label{fig:4(a)}
  \end{subfigure}
  \hfill
  \begin{subfigure}[b]{0.9\textwidth}
    \includegraphics[width=\textwidth]{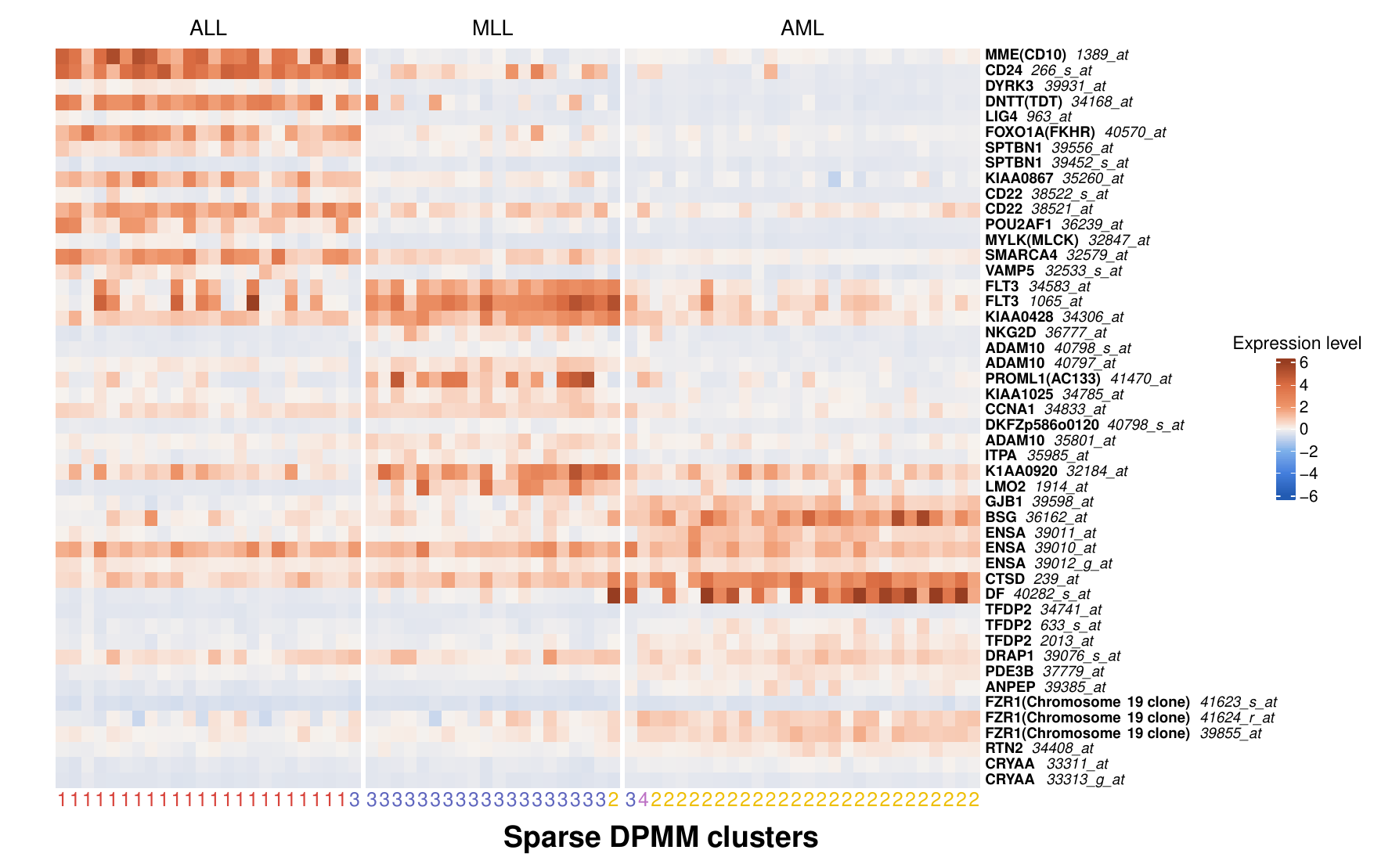}
    \caption{Heat map of clusters based on sub-type specific expressed genes}
    \label{fig:4(b)}
  \end{subfigure}
  
  \caption{\ref{fig:4(a)}) PCA projection on the first $2$ principal components for Sparse DPMM clusters obtained using weaker hyper-prarameters; \ref{fig:4(b)}) Heatmap showing comparison between labelled sub-types and Sparse DPMM (weaker) cluster estimates with respect to specifically expressed genes for the labelled sub-types \citep{armstrong2002mll} for all the $72$ samples (genes in bold, probe-id in italics)}
\end{figure}

We applied the proposed clustering framework on the gene-expression dataset from leukemia sub-type study by \citet{armstrong2002mll}, previously used as a benchmark for comparative analysis of clustering algorithms \citep{de2008clustering}. \citet{armstrong2002mll} showed that leukemia with rearrangement of the MLL gene (mixed lineage leukemia gene, now renamed KMT2A) should be classified as a distinct clinical entity rather than common acute lymphoblastic leukemia. The chromosomal rearrangement of MLL creates fusion genes with various partner genes that influence lineage commitment \citep{krivtsov2007mll}. For example, certain MLL rearrangements result in overexpression of \textbf{HOXA9} and \textbf{PRG1} genes, which are characteristic markers of acute myeloid leukemia (AML) rather than ALL \citep{armstrong2002mll}. Recent studies have further revealed the lineage plasticity of MLL-rearranged leukemia, showing dynamic transitions between lymphoid and myeloid phenotypes \citep{chen2022single, janssens2024mll}.\\
Working on the Affymetrix data from \citet{armstrong2002mll}, \citet{de2008clustering} provides a filtered set of $d = 2194$ expressed genes for clustering $N = 72$ leukemia samples into ALL-MLL-AML sub-types (Fig \ref{fig:3}(a)). For a strong choice of prior hyper-parameters, we obtained $K_{post} = 3$ with ARI score of $0.92$ (Figure \ref{fig:3}(b)).
Following our empirical Bayes approach, we estimated $a_0=b_0=28.9076$ which corresponded towards lower cluster overlap from $KL_{mean}$ values (details in Supplementary Section S6:2). Along with $k_0 = N+1 = 73$, we applied Sparse DPMM for random initial probability allocation values and chose the best performing model based on VLL. To check the sensitivity of Sparse DPMM with a weaker prior hyper-parameter, we used $a_0=b_0=10$ instead of $28.9076$ and clustered the same dataset. For this particular choice, we obtained $K_{post}=4$ with ARI score $0.86$. (Figure \ref{fig:4(a)}).
Comparing with known sub-types in Figure \ref{fig:3}(a), Sparse DPMM with weaker hyper-parameters identifies an additional $4^{th}$ cluster with one sample replacing a labelled AML sub-type (Figure \ref{fig:4(a)}). Analogous to \citep{armstrong2002mll}, we plot a heat-map based on the available genes \citep{de2008clustering} that are specifically expressed for the known sub-types to compare the lineage of the $4^{th}$ estimated cluster (Figure \ref{fig:4(b)}). The heat-map reveals a mixed expression pattern for the 4th cluster. The cluster shows elevated expression of genes characteristic of ALL sub-types (\textbf{CD22} and \textbf{CD24}), suggesting lymphoblastic features. It also displays moderate expression of genes typically overexpressed in MLL sub-types (\textbf{PROML1}, \textbf{FLT3} and \textbf{ADAM10}). In contrast, genes predominantly expressed in AML sub-types (\textbf{DF}, \textbf{CTSD} and \textbf{BSG}) are not dominantly expressed in this cluster. However, this expression pattern is not clear-cut, with the 4th cluster displaying characteristics intermediate between ALL and MLL rather than fitting cleanly into a single category. This ambiguous profile is consistent with the documented lineage plasticity of MLL-rearranged leukemias, which can undergo dynamic transitions between lymphoid and myeloid phenotypes during disease progression and treatment \citep{chen2022single, janssens2024mll}. The identification of this 4th cluster by Sparse DPMM thus appears to capture this biological reality of MLL heterogeneity rather than representing a technical artifact.

\subsubsection{Performance benchmarking with current state of the art clustering techniques}
For high-dimensional clustering analysis, there are certain popular choice of working algorithms with varying implementation techniques. Methods like DBSCAN \citep{ester1996density} and HDBSCAN \citep{campello2013density} are based on density of the data distributed over dimensional space, where data points in the low-density region are identified as noise/outliers and form clusters using core points. Shared nearest neighbours based method \citep{ertoz2002new} works by grouping data points with higher similarity of shared neighbours across a k-nearest neighbour graph. \citet{bouveyron2007high} proposed a high dimensional model based clustering that implements dimensionality reduction for each cluster with parsimonious covariance structure and estimate the Gaussian mixture model parameters using EM iterations. Another class of popular methods include graph based algorithms, like Leiden (an upgrade of Louvain, \citet{traag2019louvain}) that constructs a k-nearest neighbour graph from the data and detects communities (clusters) by maximizing graph modularity. Classical $K$-means \citep{mcqueen1967some}, on the other hand, is a partitioning based method that minimizes the within-cluster variance by iteratively assigning data points to their nearest centroids.
\begin{figure}[!ht]
    \centering
    \includegraphics[width=0.75\textwidth]{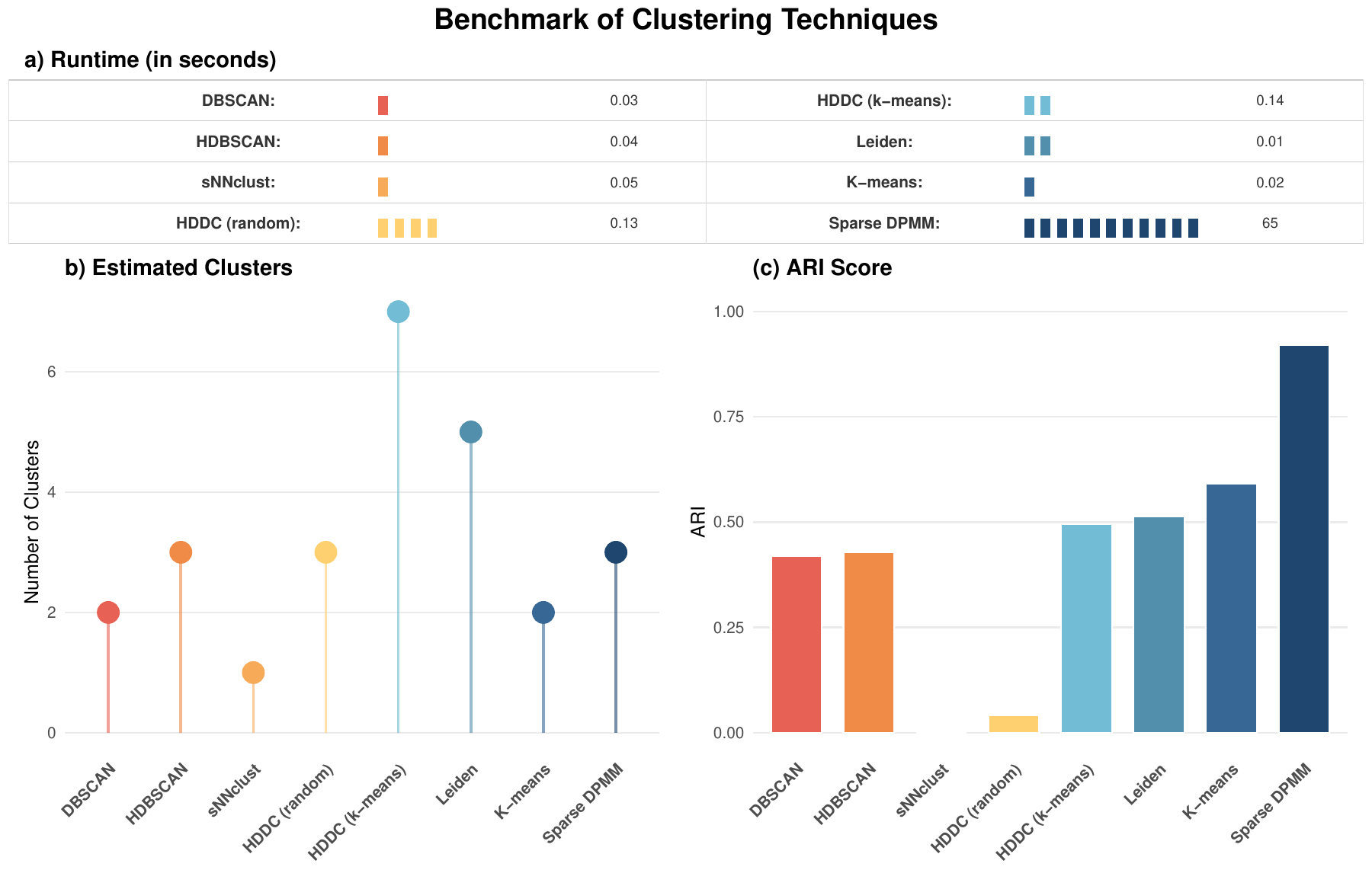}
    \caption{Performance benchmarking of popular clustering methods against Sparse DPMM; based on $a)$ implementation run-time in seconds (iterations are indicated with mini-bars), $b)$ estimated number of clusters and $c)$ corresponding ARI scores}
    \label{fig:5}
\end{figure}
For performance on the Leukemia dataset, we used the R packages implementing those methods and hyperparameter choice was based on each method default recommendations (see details in Supplementary Material S8). Based on corresponding optimal parameters, these methods are compared with Sparse DPMM implementation with empirical Bayes hyper-parameters, and we validated their performances with run-time (in seconds), estimated cluster numbers and ajusted Rand index (ARI). Due to diversity in methodology of these techniques, we avoided any internal metric for comparison across methods and focused on external validation metrics. Figure \ref{fig:5} shows that Sparse DPMM is significantly slower ($\sim 5.91$ seconds per iteration) in comparison to alternative methods. However, the performance metrics based on estimated number of clusters and ARI scores validate the (relative) superiority of Sparse DPMM.

\section{Discussion}\label{sec:4}
We developed a Bayesian non-parametric mixture model using collapsed variational inference (VI) to perform unsupervised clustering. Our approach hierarchically incorporates the Dirichlet process (DP) concentration parameter $\alpha$ as a variable in the DP mixture model (DPMM) with Gaussian kernels. Based on the importance of adaptive estimation of $\alpha$ \citep{ascolani2023clustering}, the proposed framework enhances the consistency of collapsed VI model developed by \citet{kurihara2007collapsed}. We also illustrated the effect of prior for covariance matrix, and concluded that a cluster-specific Sparse DPMM is the best choice. Implementing our theoretical work into an R-package \texttt{vimixr}, we provide a probabilistic tool to perform exploratory analysis of unstructured and high-dimensional biological data with significant potential applicability.\\
The Sparse DPMM provides significantly faster convergence than an MCMC slice sampling approach \citep{hejblum2019sequential}. Comparing with conventional clustering techniques, the iterative nature of Sparse DPMM, added to the model structure complexity, leads to a slower convergence. However, it provides better clustering estimates with empirical Bayes hyper-parameters using gene expression data for leukemia sub-types \citep{armstrong2002mll}. Even with weaker hyper-parameters, Sparse DPMM estimates are consistent with gene expression profiles, hence providing a biologically meaningful sub-cluster of the known sub-types. This is particularly reflected by the PCA plot in Figure \ref{fig:4(a)} where the $4^{th}$ cluster has more numerical proximity towards AML subtype, but the marker genes heatmap in Figure \ref{fig:4(b)} represents more biological proximity towards ALL subtype.\\
A natural extension of Sparse DPMM is to generalize beyond Gaussian kernels for $G_0$, which becomes eminent for real-world data. For instance, overdispersed count data is better modelled using Negative Binomial distributions, thus avoiding model misspecification for clustering such datasets like gene expression data \citep{anders2010differential}. Even under such misspecified conditions, Sparse DPMM is able to retrieve malignant cell types across HSC-myeloid differentiation axis based on gene subclones from single cell transcriptomics (details in Supplementary S9). Another aspect is to introduce probabilistic feature selection variables that could be updated simultaneously \citep{tadesse2005bayesian} or guide the clustering process itself \citep{rouanet2024bayesian}. This can provide insights on the features defining the estimated clusters and prove beneficial for high-dimensional data sets.\\
As an optimization technique, VI algorithms heavily depend on the initial choice of latent variables (latent allocation vectors $\{z_n\}$ in our case) for a given data. We advocate random initialization for these parameters, and the choice is based on variational log likelihood (VLL) values. For VI implementation, alternative approaches \citep{blei2017variational} can be explored for further refinement. Based on the analytical properties of VLL, empirical Bayes estimates for the $G_0$ hyper-parameters prove to be efficient for robust estimation. Due to their implied significance over effective dimensions and sample size, it becomes important to consider hierarchical hyper-priors for the hyper-parameter $a_0=b_0$. However, the conjugate prior for both unknown shape and rate parameters of a Gamma distribution does not have a closed form \citep{miller1980bayesian}. So, incorporating an $a_0=b_0$ hyper-prior would add another layer of analytical complexity, along with increased computational cost due to additional model parameters.\\
The exchangeability of mixture components in DPMM introduces label switching between atoms as they are ordered sequentially through the stick-breaking construction \citep{papaspiliopoulos2008retrospective}. Following \citet{kurihara2007collapsed}, we addressed the issue by ordering the latent allocation probability updates such that the clusters are arranged in decreasing proportions. Alternatively, \citet{fruhwirth2011dealing} shows identification methods for finite mixture model, that can be extrapolated to non-parametric mixtures.

\section*{Acknowledgments}
This work was supported by the University of Bordeaux’s Digital Public Health Graduate School, financed by the PIA 3 through the Agence Nationale
de la Recherche (Ref: 17-EURE-0019), by the Programme et Equipement Prioritaire de Recheche Santé Numérique (PEPR SN) project SMATCH (Ref:
22-PESN-0003). Computer time for this study was provided by the computing facilities MCIA (Mésocentre de Calcul Intensif Aquitain) of the Université de Bordeaux and of the Université de Pau et des Pays de l’Adour.

\bibliographystyle{apalike}
\bibliography{biomsample.bib}

\appendix

\section{}
\subsection{Software and Data availability}
Software in the form of R package \textbf{vimixr} with complete documentation is available on CRAN (\texttt{https://cran.r-project.org/web/packages/vimixr/index.html}). The Leukemia sub-type gene expression data is publicly available online at\\ \texttt{https://schlieplab.org/Static/Supplements/CompCancer/Affymetrix/armstrong-2002-v2/}. The package implementation is provided in the Zenodo repository\\ \texttt{https://doi.org/10.5281/zenodo.18405687}, along with metadata and results.

\end{document}